# Systematic Extraction of Spectroscopic Factors from $^{12}$C(d,p)$^{13}$C and $^{13}$C(p,d)$^{12}$C Reactions

X. Liu, M.A. Famiano, W.G. Lynch, M.B. Tsang,
National Superconducting Cyclotron Laboratory and Department of Physics and Astronomy, Michigan State University, East Lansing, MI 48824

J.A. Tostevin
Department of Physics, School of Electronics and Physical Sciences,
University of Surrey, Guildford, Surrey GU2 7XH, United Kingdom

**Abstract**

Existing measurements of the angular distributions of the ground-state to ground-state transitions of the $^{12}$C(d,p)$^{13}$C and $^{13}$C(p,d)$^{12}$C neutron-transfer reactions have been analyzed systematically using the Johnson-Soper adiabatic and distorted-wave theories. When using a consistent set of physical inputs the deduced spectroscopic factors are consistent to within 20% for incident deuteron energies from 6 to 60 MeV. By contrast, original analyses of many of these data quoted spectroscopic factors that differed by up to a factor of five. The present analysis provides an important reference point from which to assess the requirements of future spectroscopic analyses of transfer reactions measured in inverse kinematics using rare nuclei.

The ordering and occupancies of the single nucleon orbits influences the energies and angular momenta of low-lying quantum states of nuclei [1,2], as well as their decays and production rates in nuclear experiments and stellar environments [3]. Single-nucleon transfer reactions, such as d+A $\Rightarrow$ p+B and p+B$\Rightarrow$ d+A, probe the spectroscopy [1-6] of the companion target or target-like nuclei, providing occupancies or spectroscopic factors for their single-particle orbitals. The overlap integral between the wave function of one state in nucleus A and another in B defines the theoretical spectroscopic factor for transfer between these states [5,6]. The ratio of the corresponding measured cross-section divided by the cross section calculated for unit spectroscopic factor provides its experimental counterpart.

The ordering and occupancies of the valence orbitals for unstable nuclei far from the valley of beta stability may differ from those of stable nuclei, leading to novel and surprising properties for the corresponding unstable nuclear states [7,8]. Single-nucleon transfer and knockout experiments in inverse kinematics with rare nuclei as projectiles, provide the optimal way to study neutron- and proton-rich nuclei and their single-particle states [3,9-12]. Rare isotope beam intensities remain very much less than those for stable beams. Since the history and experience of transfer reaction-based spectroscopy using rare isotopes is much shorter, it is critical to understand the limitations of reaction theory by selectively reexamining the consistency of the analyses of more precise measurements made with intense light-ion beams, and to develop strategies to overcome them.

Large uncertainties can be associated with the extraction of spectroscopic factors from transfer reactions. In a systematic compilation of spectroscopic factors for sd-shell nuclei, Endt [13] noted that very different values of the spectroscopic factors arise from different analyses and/or experiments. By examining a large amount of data, and using consistency checks when available, Endt compiled a list of "best" spectroscopic factor values for the sd-shell. This analysis, performed in 1977, does not provide the systematic uncertainties associated with the method. Nor did it take advantage of improved model calculations that include deuteron break-up effects.

The sensitivity of calculations to the optical model potentials assumed in the entrance and exit channels constitutes the most significant problem in the extraction of spectroscopic factors from DWBA analyses of transfer reactions [5,14-17]. The deduced spectroscopic factors are dependent on these choices; thus, many have advocated measuring elastic scattering data at the entrance and exit energies to fix the optical model parameters. However, such data are neither available nor easily obtained at the desired energy for most reactions. The analysis proposed here minimizes the need for such data and shows that superior results can be achieved by the consistent use of reasonable theoretical inputs.

Many normal kinematics transfer reaction data have been accumulated. Theoretical advances have provided both global, phenomenological nucleon optical potentials [18-23] and potentials based on microscopic methods, such as the Jeukenne-Lejeune- Mahaux (JLM) effective nucleon-nucleon interaction [24,25]. To assess the systematic uncertainties associated with the extraction of spectroscopic factors, we choose a reaction measured over a large range of incident energies using different detection systems. Specifically, we analyze the angular distributions of the differential cross-section of the $^{12}$C(d,p)$^{13}$C(g.s.) reaction and of its inverse, $^{13}$C(p,d)$^{12}$C(g.s.).

There are published angular distributions for the $^{12}$C(d,p) reaction at incident deuteron energies from 0.4 to 56 MeV [14,15,26-44], and for the inverse reaction from 35 to 70 MeV [16,17,45,46]. Until now, spectroscopic factors have been extracted from only a subset of these experiments. The associated analyses relied mainly on distorted wave Born approximation (DWBA) calculations without a consistent choice of input parameters. Figure 1 shows the published $^{12}$C(d,p)$^{13}$C(g.s.) (closed points) and $^{13}$C(p,d)$^{12}$C(g.s.) (open points) spectroscopic factors as a function of the equivalent incident deuteron energy. The values fluctuate from 0.3 to 1.4 with no evident correlation with incident energy. In some experiments [14,15,16,17], multiple values were deduced from different optical model parameter sets; the higher values are shown for those cases as squares in Figure 1. The dashed line shows the theoretical prediction (0.62) of the Cohen and Kurath shell model calculation [47]. The scatter illustrates the problem of extracting consistently a meaningful empirical spectroscopic factor.

In the present analysis we consider the $^{12}$C(d,p) angular distributions, shown in Figure 2, that have been measured at incident energies ranging from 7 MeV to 56 MeV. Each is displaced by factors of 10 from the neighboring distributions for ease of presentation, the overall normalization factor being unity for the 19.6 MeV angular distribution. As is normal, the spectroscopic factor is extracted by fitting the theory to the data at the first peak in the angular distribution, since the backward angle data are more sensitive to the effects of inelastic couplings and other higher-order effects. To be consistent, the spectroscopic factors are extracted by minimizing $\chi^2$, including only angular points that are (i) within 30% of the predicted maximum yield and (ii) at $\theta_{cm}<$ 30°. A similar analysis is also performed on data from the inverse reaction of $^{13}$C(p,d)$^{12}$C(g.s.) for proton incident energies ranging from 35 to 65 MeV.

Our calculations analyze all data sets in a consistent way and use a modified version of the code TWOFNR [48]. All calculations make the local energy approximation (LEA) for finite range effects [49] using the zero-range strength ($D_0$) and range ($\beta$) parameters of the Reid Soft core $^3S_1$-$^3D_1$ neutron-proton interaction [50]. The $1p_{1/2}$ neutron binding potential had radius parameter 1.25 fm and diffuseness 0.65 fm. Non-locality corrections [51] with range parameters of 0.85 and 0.54 are included in the proton and deuteron channels, respectively.

We calculate the transfer cross sections within the Johnson-Soper (JS) adiabatic approximation to the neutron, proton and target three-body system [52], which includes the effects of break up of the deuteron in the field of the target, and requires *only* a specification of the nucleon-target interactions. The exact adiabatic three-body model (d,p) and (p,d) transfer reaction amplitudes require knowledge of the adiabatic three-body wave function only at small neutron-proton separations. There, the adiabatic distorting potential governing the center-of-mass motion of the deuteron is well described by the sum of the neutron- and proton-target optical potentials [52]. It is important to stress that this adiabatic distorting potential generates the three-body wave function in that limited region of configuration space needed to evaluate the transfer amplitude, and it does not describe deuteron elastic scattering at the beam energy.

We first perform (d,p) calculations where both the exit channel proton potential and the entrance channel JS adiabatic potential use the JLM nucleon-target optical potentials [53]. These are calculated by folding the density-dependent JLM nucleon-nucleon effective interaction [24,25], assumed to have a Gaussian form factor of range 1 fm [54], with the assumed target matter density in the mid-point local-density approximation [54]. The matter density distributions for both $^{12}$C and $^{13}$C are evaluated assuming the modified oscillator density parameters ($\alpha$=1.247, a=1.649 fm for $^{12}$C; $\alpha$=1.403, a=1.635 fm for $^{13}$C ) compiled in ref. [55]. The corresponding root-mean-square (rms) charge radii are 2.46 fm and 2.44 fm for $^{12}$C and $^{13}$C, respectively. The real and imaginary parts of the calculated nucleon optical potentials were scaled by multiplicative factors $\lambda_v$=1.0 and $\lambda_w$=0.8, respectively, obtained from a systematic study of light nuclei [53].

The calculated angular distributions normalized by the extracted spectroscopic factors are shown as solid lines in Figure 2. The associated spectroscopic factors, shown at the bottom of Figure 3, include reanalyzes of the data in Figure 1 (closed circles) [14,29,31,32,35,38,39] and of additional data sets (closed squares) [30,33-36]. Above 35 MeV, we supplement the limited (d,p) data with $^{13}$C(p,d) data [16,17,45,46] denoted by the open symbols in Figure 3. We did not analyze data at $E_d$=28 [37] and 56 MeV [15], and at $E_p$=65 MeV [17] because those angular distributions did not include the first peak. The spectroscopic factors deduced for $E_d$= 6-60 MeV provide an average deduced spectroscopic factor of 0.63± 0.11. This energy range spans the optimum angular and linear momentum matching conditions for (d,p) and (p,d) transfer reaction studies. In contrast, the published values in Figure 1 vary from 0.3 to 1.5. Our consistent, theoretically motivated analyses reduce the fluctuations substantially.

To assess the stability of the above adiabatic three-body model calculations, we have repeated these analyses while replacing the JLM nucleon optical potentials everywhere by the Chapel Hill (CH) [20] global potential set. The spectroscopic factors are shown in the center of Figure 3. Overall, the values are quite similar, but are consistently somewhat higher and with greater departures from the JLM values for energies above 35 MeV. Below 35 MeV, the average spectroscopic factor is 0.73±0.1, within the uncertainties obtained in the analysis with JLM potentials. It should be noted that light nuclei were not included in the database for the CH potential evaluation. A consistent use of alternative global nucleon potentials, such as that of Bechetti and Greenless [21], was found to lead to very similar results to those of the CH set.

For a final comparison, we also analyzed the full data set within the DWBA formalism, neglecting the role of deuteron break-up channels. To remove energy-dependent optical potential ambiguity, we used the CH and Daehnick [22] global potentials for the proton and deuteron channels, respectively. The calculated angular distributions normalized by the spectroscopic factors are shown by the dashed curves in Figure 2 and the deduced spectroscopic factors are plotted at the top of Figure 3. The symbol convention is the same as for the JLM and CH cases. The extracted values are far more consistent than the published values shown in Figure 1. The average value is 0.79±0.08 from those data below 35 MeV. For the higher energies, the values are much larger. Comparisons with the JS adiabatic calculations suggest that neglect of the break-up channel within the DWBA may be a significant contributing factor at these higher energies.

In summary, published analyses of angular distributions for the $^{12}C(d,p)^{13}C(g.s.)$ and $^{13}C(p,d)^{12}C(g.s.)$ reactions display considerable variations in the extracted spectroscopic factors. Using a consistent optical potential parameterizations and fitting the first maximum in the angular distributions provides spectroscopic factors that are consistent to within 20% over a wide range of energy. The use of global optical potentials or the JLM potential parameterization results in a similar behavior for the spectroscopic factors; the absolute value depends somewhat on the potential choice used.

The global parameterization of the optical model potentials requires the minimum input parameter choice and may be advantageous where it is not important to extract the absolute spectroscopic factors. These global parameterizations, however, are fitted only to data from stable nuclei. In neutron and proton-rich regions of the nuclear chart, the use of microscopic optical potentials like the JLM, but folded with Hartree-Fock matter densities for the relevant nuclei may offer a more realistic alternative. The current analysis of $^{12}C(d,p)^{13}C$ and $^{13}C(p,d)^{12}C$ reactions provide reference points in the p-shell to which relative spectroscopic factors can be measured. The input parameters, summarized in Table 1, should be broadly applicable to other systems. Similar studies of other systems are needed to guide the broadly-based programs investigating the single particle structure of rare isotopes with the operation of current and future intense rare isotope facilities.

This work is supported by the National Science Foundation under Grant No. PHY-01-10253 and by the UK Engineering and Physical Sciences Research Council through Grant No. GR/M82141.

Table 1: Summary of the input parameters used in TWOFNR

|  | DWBA | Adiabatic CH | JLM |
|---|---|---|---|
| Proton potential | Chapel-Hill [20] | Chapel-Hill [20] | JLM [24,25] |
| Deuteron potential | Daehnick [22] | Adiabatic [52] from CH | Adiabatic [52] from JLM |
| Target densities |  |  | Modified Harmonic Oscillator density [55] $\alpha$=1.247, a=1.649 fm $^{12}$C $\alpha$=1.403, a=1.635 fm $^{13}$C |
| n-binding potential | Woods-Saxon, $r_0$=1.25, a=0.65, depth adjusted, no spin-orbit | Woods-Saxon, $r_0$=1.25, a=0.65, depth adjusted, no spin-orbit | Woods-Saxon, $r_0$=1.25, a=0.65, depth adjusted, no spin-orbit |
| Finite range | Yes | Yes | Yes |
| Hulthen finite range factor [50] | 0.7457 | 0.7457 | 0.7457 |
| Vertex constant $D_0^2$, [50] | 15006.25 | 15006.25 | 15006.25 |
| JLM potential scaling $\lambda$ | N/A | N/A | $\lambda_v$=1.0 and $\lambda_w$=0.8 [53] |
| Non-Locality | yes | yes | yes |
| Non-Locality potentials | p 0.85; n N/A; d 0.54 | p 0.85; n N/A; d 0.54 | p 0.85; n N/A; d 0.54 |

**References:**

[1] N. Austern, Direct Nuclear Reaction Theories, Wiley, New York, 1970.

[2] G.R. Satchler, Direct Nuclear Reactions, Oxford University Press, Oxford, 1983.

[3] W.N. Catford, Nucl. Phys. **A 701**, 1c (2002).

[4] S.T. Butler, Proc. Roy. Soc. (London) **A208**, 559 (1951)

[5] M.H. Macfarlane and J.B. French, Rev. Mod. Phys. **32**, 567 (1960).

[6] M.H. Macfarlane and J.P. Schiffer , Nuclear Spectroscopy and Reactions, Vol. B, page 170-194, Academic Press, New York and London, 1974.

[7] P.G. Hansen, A.S. Jensen and B. Jonson, Ann. Rev. Nucl. Part. Sci. **45,** 505 (1995).

[8] V. Maddalena et al., Phys. Rev. C 63, 024613 (2001).

[9] P.G. Hansen and B.M. Sherrill, Nucl. Phys. **A 693,** 133 (2001).

[10] P.G. Hansen and J.A. Tostevin, Ann. Rev. Nucl. Part. Sci. **53,** 219 (2003).

[11] S. Fortier, S. Pita, J.S. Winfield, et al., Phys. Lett. B **461**, 22 (1999).

[12] J.S. Winfield, S. Fortier, W.N. Catford, et al., Nucl. Phys. A **683,** 48 (2001).

[13] P.M. Endt, Atomic Data and Nuclear Data Tables **19,** 23 (1977).

[14] S.E. Darden, S. Sen, H.R. Hiddleston, J.A. Aymar and W.A. Toh, Nucl. Phys. **A 208**, 77 (1973).

[15] K. Hatanaka,  N. Matsuoka, T. Saito, et al.,  Nucl. Phys. **A 419**, 530 (1984).

[16] J.R. Campbell, W.R. Falk, N.E. Davison, J. Knudson and R. Aryaeinejad, Nucl. Phys. **A 470**, 349 (1987).

[17] K. Hosono, M. Kondo and T. Saito, Nucl. Phys.**A 343,** 234 (1980).

[18] C.M. Perey , F.G. Perey Atomic Data and Nuclear Data Tables **17**, p6 (1976).

[19] J.J.H. Menet, E.E. Gross, J.J. Malanify and A. Zucker, Phys. Rev. **C 4,** 1114 (1971).

[20] R.L.Varner, W.J. Thompson, T.L. McAbee, E.J. Ludwig and T.B. Clegg, Phys. Rep. **201,** 57 (1991).

[21] F.D. Bechetti, Jr., G.W. Greenless, Phys. Rev. **182**, 1190 (1969).

[22] W.W. Daehnick, J.D. Childs and Z. Vrcelj, Phys. Rev. **C 21**, 2253 (1980).

[23] J.M. Lohr and W. Haeberli Nucl.Phys. **A 232**,  381 (1974).

[24] J.-P. Jeukenne, A.  Lejeune and C. Mahaux, Phys.Rev. **C 15**, 10  (1977).

[25] J.-P. Jeukenne,  A. Lejeune, C. Mahaux, Phys. Rev. **C 16**, 80 (1977).

[26] A. Gallmann, P. Fintz, P.E. Hodgson Nucl. Phys. **82**, 161 (1966).

[27] T.W. Bonner, J.T. Eisinger, A.A. Kraus, Jr., J.B. Marion, Phys. Rev. **101**, 209 (1956).

[28] H. Guratzsch, G. Hofmann, H. Muller, G. Stiller, Nucl. Phys. **A 129**, 405 (1969).

[29] N.I. Zaika, et al., Soviet Phys. JETP **12**, 1 (1961).

[30] D. Robson, Nucl. Phys. **22**, 34 (1961).

[31] U. Schmidt-Rohr, R.Stock, and P.Turek, Nuc. Phys. **A 53**, 77 (1964).

[32] J. Lang, J.Liechti, R.Muller, et al, Nuc. Phys. , 77 (1988).

[33] E.W. Hamburger, Phys. Rev. **123**, 619 (1961).

[34] J.N. McGruer, Phys. Rev. **100**, 235 (1955).

[35] S. Morita, N. Kawai, N. Takano, Y. Goto, R. Hanada, Y. Nakajima, S. Takemoto, and Y. Taegashi, J. Phys. Soc. Japan, **15**, 550 (1960).

[36] R. van Dantzig, L.A. CH. Koerts, Nuc. Phys. **48**, 177 (1963).

[37] R.J. Slobodrian, Phys. Rev. **126**, 1059 (1962).

[38] H. Ohnuma, N. Hoshino, O. Mikoshiba, et al., Nucl. Phys. **A 448**, 205 (1986).

[39] W. Fetscher, K. Sattler, E. Seibt, R. Staudt and Ch.Weddigen. Proc. Third Inter. Symp. on Polarization Phenomena in Nuclear Reactions. Eds. H.H. Barschall and W. Haeberli (University of Wisconsin Press, Madison, 1971), p772.

[40] J.P. Schiffer, G.C. Morrison, R.H. Siemssen and B. Zeidman, Phys. Rev. **164,** 1274 (1967).

[41] R.V. Poore, P.E. Shearin, D.R. Tilley and R.M. Williamson, Nucl. Phys. **A 92,** 97 (1967).

[42] N.E. Davison, P. Fintz and A. Gallmann, Nucl. Phys. **A 220**, 166 (1974).

[43] Graeme D. Putt, Nucl. Phys. **A 161,** 547 (1971).

[44] J.W. Leonard, D.O. Wells, Nucl. Phys. **A 153,** 657 (1970).

[45] H. Toyokawa, H. Ohnuma, Y. Tajima, et al., Phys. Rev. **C 51**, 2592 (1995).

[46] H. Taketani, J. Muto, H. Yamaguchi and J. Kokame, Phys. Lett. **B 27**, 625 (1968).

[47] S. Cohen and D. Kurath, Nucl. Phys. **A 101**, 1 (1967).

[48] M. Igarashi, et al, Computer Program TWOFNR (Surrey University version).

[49] P.J.A. Buttle and L.J.B. Goldfarb, Proc. Phys. Soc. London, **83**, 701 (1964).

[50] L.D. Knutson, J.A. Thomson and H.O. Meyer, Nucl. Phys. **A 241**, 36 (1975).

[51] F. Perey and B. Buck, Nucl. Phys. **32,** 353 (1962).

[52] R.C. Johnson and P.J.R. Soper, Phys. Rev. **C 1,** 976 (1970).

[53] J. S. Petler, M. S. Islam, R. W. Finlay, and F. S. Dietrich, Phys. Rev. **C 32**, 673 (1985).

[54] S. Mellema, R. W. Finlay, F. S. Dietrich, and F. Petrovich, Phys. Rev. **C 28**, 2267 (1983).

[55] C.W. De Jager, H. De Vries and C. De Vries, Atomic Data and Nuclear Data Tables **14**, 479 (1974).

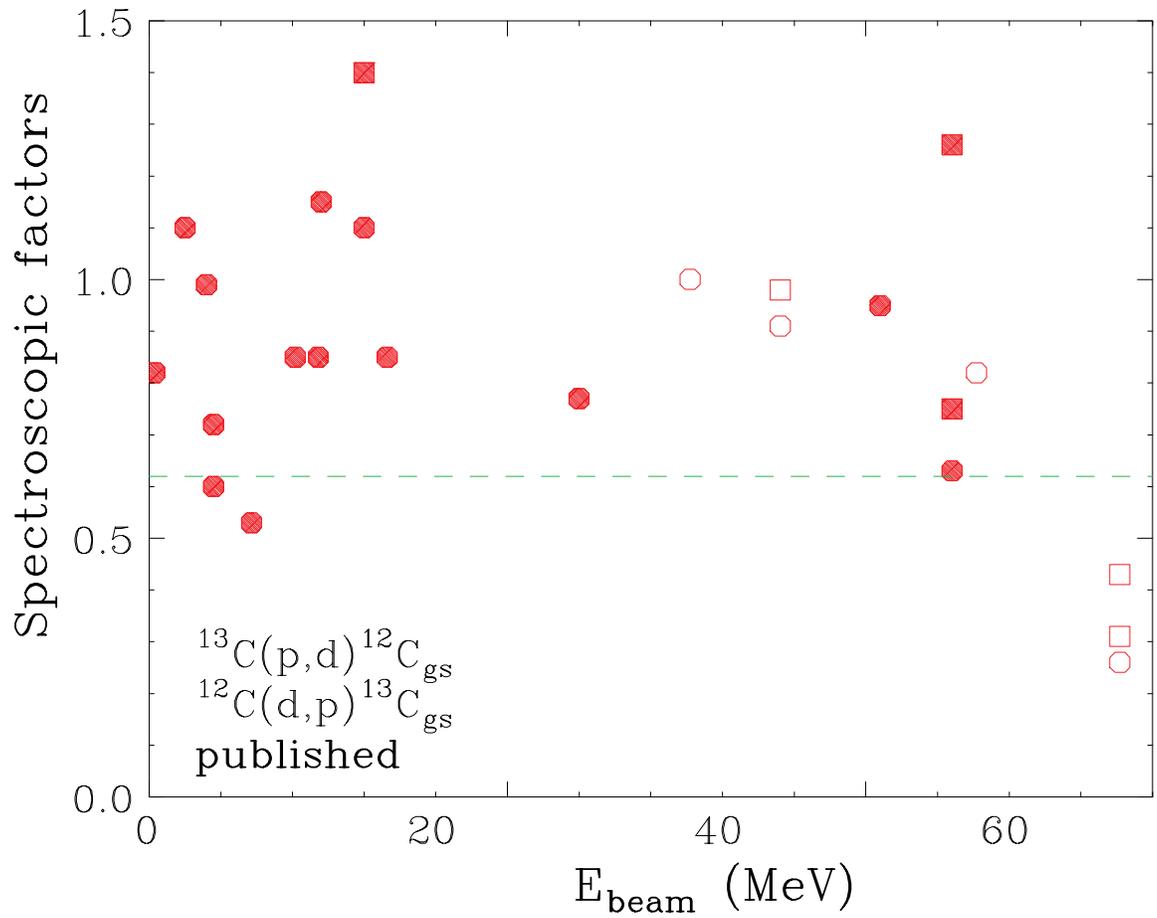

**Figure 1**

Spectroscopic factors for $^{12}C(d,p)^{13}C(gs)$ and $^{13}C(p,d)^{12}C(gs)$ reactions extracted from the literature [14-17, 26, 27, 29, 31-33, 35, 39, 43, 45, 46].

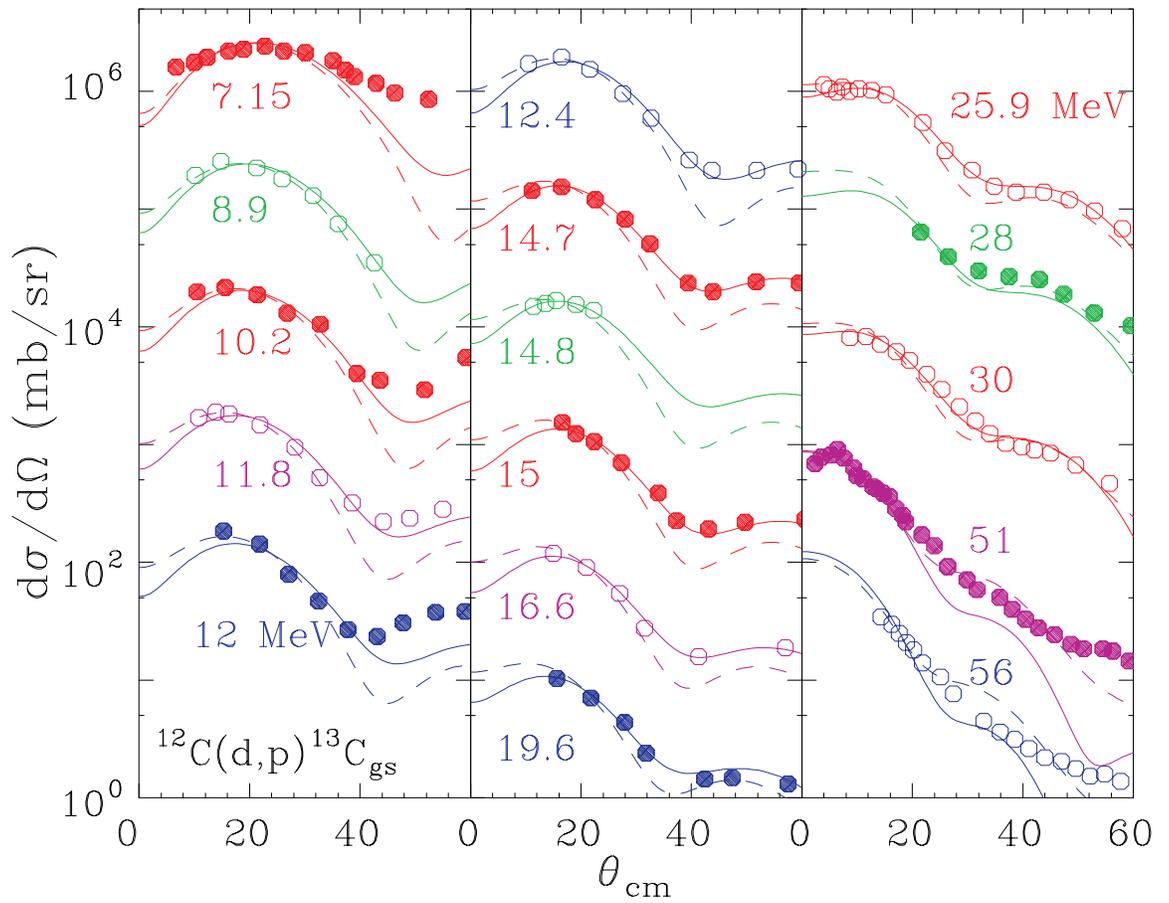

**Figure 2**

Published angular distributions for $^{12}$C(d,p)$^{13}$C reactions.

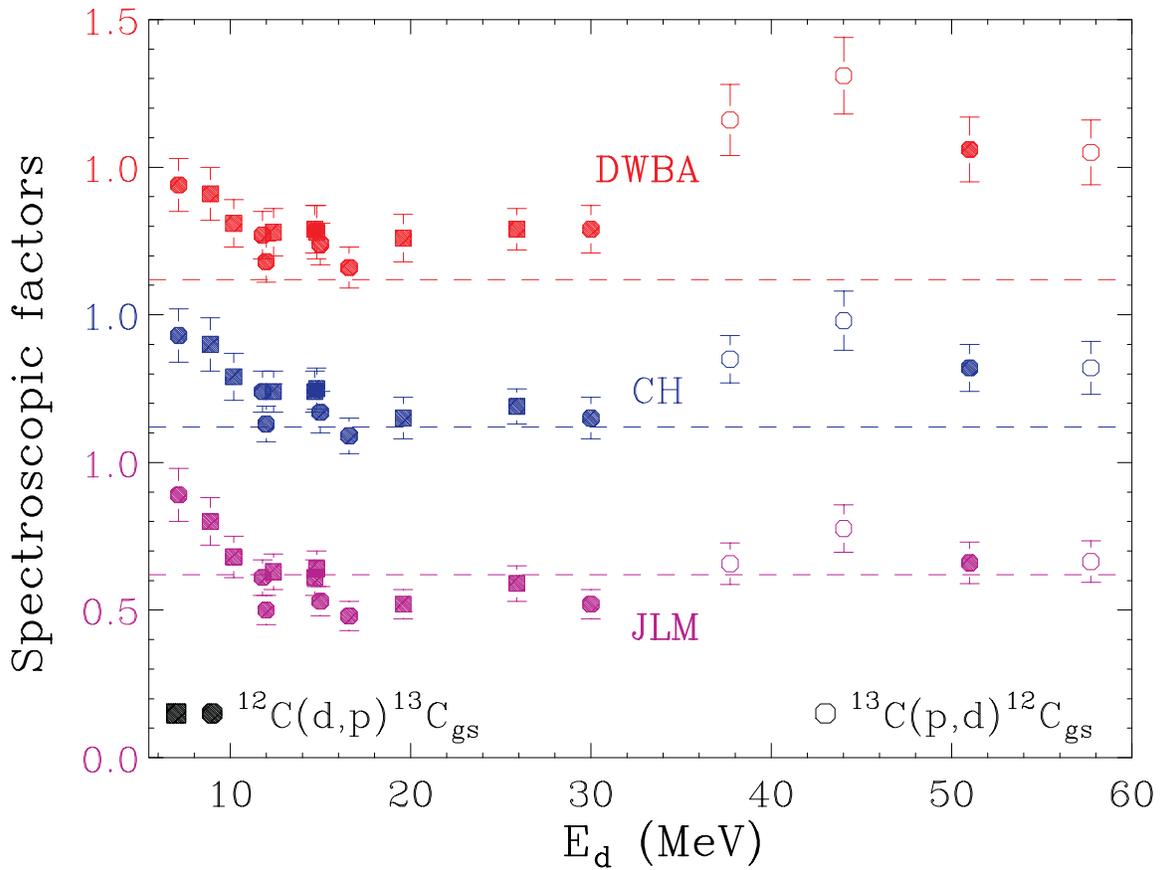

**Figure 3**
Extracted spectroscopic factors in the present work for $^{12}C(d,p)^{13}C$ and $^{13}C(p,d)^{12}C$ reactions. The dashed lines represent the shell model prediction of Cohan and Kurath [47] of 0.62.. Results from three different analyses using the parameters summarized in Table I are shown. See text for detail explanation.